\documentclass[epjST]{svjour}

\usepackage{graphicx,bbm,setspace,amssymb,amsfonts,amsmath,mathtools,mathrsfs,stmaryrd,bm,bm,etoolbox,comment,hyperref,url,color,enumitem,algorithm,booktabs,microtype,nicefrac,lingmacros,nccmath,tree-dvips,tikz-cd,xurl,soul}

\DeclareMathOperator*{\argmin}{argmin}
\DeclareMathOperator*{\argmax}{argmax}

\usepackage{graphicx}
\usepackage{subcaption}
\usepackage{graphics}
\begin{document}
\title{Estimating a continuously varying offset between multivariate time series with application to COVID-19 in the United States}
\author{Nick James\inst{1} \and Max Menzies \inst{2}\fnmsep\thanks{\email{max.menzies@alumni.harvard.edu}}}
\institute{School of Mathematics and Statistics, University of Melbourne, Victoria 3010, Australia \and Beijing Institute of Mathematical Sciences and Applications, Tsinghua University, Beijing 101408, China}
\abstract{
This paper introduces new methods to track the offset between two multivariate time series on a continuous basis. We then apply this framework to COVID-19 counts on a state-by-state basis in the United States to determine the progression from cases to deaths as a function of time. Across multiple approaches, we reveal an ``up-down-up'' pattern in the estimated offset between reported cases and deaths as the pandemic progresses. This analysis could be used to predict imminent increased load on a healthcare system and aid the allocation of additional resources in advance.
} 
\maketitle
\section{Introduction}
\label{sec:intro}

Understanding the trajectories of and relationships between COVID-19 case and death counts assists governments in anticipating and responding to the impact of the pandemic. In the United States (US) and elsewhere, high case counts have generally been closely followed by high hospital admissions, the use of costly equipment such as ICU beds and ventilators \cite{cutler2020}, and deaths. The strain on the healthcare system may be considerable and can even threaten the health of patients who are not afflicted by COVID-19 \cite{Rosenbaum2020}.

Unfortunately, the dynamics of the COVID-19 pandemic have been consistently difficult to describe and predict. Numerous factors may influence the virus' spread, including the emergence of new variants, changes in government policy and restrictions, community adherence to health recommendations, exhaustion with mitigation measures \cite{exhaustiondasilva2021}, community frustration, differing risk appetites by different population groups, and changing testing policies \cite{Pullano2020,DiBari2020,Antigenchange,Francechange}. Thus, the actual and recorded counts of COVID-19 cases have exhibited complex dynamics since the arrival of the pandemic. One of the most significant attributes to be aware of is the delay between the onset of cases and their progression to deaths. This may predict the peak of hospitalisations and provide advance warning of increased loads on the healthcare system.

This paper serves this purpose by providing an in-depth mathematical study on the estimated average offset between reported cases and deaths, investigating this as a function of time. Several clinical trials and mathematical studies have aimed to do this in isolated incidences, but this paper is the first we are aware of to develop a nonlinear dynamical framework to calculate a continuously changing time-varying offset. In Section \ref{sec:methodology}, we describe several approaches applicable to any two multivariate time series, and we report our results on the US in Section \ref{sec:results}. We supply a more in-depth discussion in Section \ref{sec:discussion}.

This paper builds on a long literature of \emph{multivariate time series analysis} and a rich literature of nonlinear dynamics applied to the COVID-19 pandemic. Existing methods of time series analysis include parametric models \cite{Hethcote2000} such as exponential  \cite{Chowell2016} or power-law models \cite{Vazquez2006} and nonparametric methods such as distance analysis \cite{Moeckel1997}, distance correlation \cite{Szkely2007,Mendes2018,Mendes2019} and network models \cite{Shang2020}. Mathematical approaches to the COVID-19 pandemic are almost too numerous to cover. First, many papers based on existing mathematical models, such as the Susceptible–Infected–Recovered (SIR) model and the (effective) reproductive ratio $R_t$ \cite{Bonifazi2021}, have been proposed and systematically collated by researchers \cite{Wynants2020,ModellingEstrada2020}. Next, nonlinear dynamics researchers have proposed several sophisticated extensions to the classical predictive SIR model, including finding analytical solutions \cite{SIRBarlow2020,SIRWeinstein2020}, modifications with additional variables \cite{SIRNg2020,SIRVyasarayani2020,SIRCadoni2020,SIRNeves2020,SIRComunian2020,Abidemi2021}, incorporation of Hamiltonian dynamics \cite{SIRBallesteros2020} or network models \cite{SIRLiu2021}, and a closer analysis of uncertainty in the SIR equations \cite{Gatto2021}. Other mathematical approaches to prediction and analysis include power-law models \cite{Manchein2020,Blasius2020,Beare2020}, forecasting models \cite{Perc2020}, fractal curves \cite{Gowrisankar2020}, Bayesian methods \cite{Manevski2020}, regression models and feature selection \cite{Gross2020,Maiti2021}, Markov chain Monte Carlo models \cite{Paul2020}, distance analysis \cite{James2021_virulence,James2020_nsm}, network models \cite{Karaivanov2020,Ge2020,Xue2020}, analyses of the dynamics of transmission and contact \cite{Saldaa2020,Danchin2021},  clustering \cite{Machado2020,jamescovideu} and many others \cite{Ngonghala2020,Cavataio2021,james2021_mobility,Nraigh2020,Glass2020,Jamesfincovid}. Finally, numerous articles have been devoted specifically to the dynamics of COVID-19 in the United States \cite{james2020covidusa}, including incorporating spatial components of the virus' spread \cite{Zhou2020_covidUS,Wang2020_spatioUS,James2021_geodesicWasserstein}. Our paper builds on this rich literature by developing a new mathematical method and a more extensive analysis of the progression of COVID-19 cases to deaths in the US than previously performed.

\section{Methodology}
\label{sec:methodology}

Our data spans 26 February 26 2020 to 25 May 2021  across $n=51$ regions (50 US states and the District of Columbia), a period of $T=454$ days. We begin here to avoid periods of sparse reporting early in the pandemic. We end here due to changes in the CDC's reporting of case data, particularly between vaccinated and unvaccinated individuals, which will be detailed in Section \ref{sec:discussion}. We order the states alphabetically and index them $i=1,...,n$. Let $x_i(t), y_i(t)$ be the multivariate time series of new daily COVID-19 cases and deaths, respectively, in each of the $n$ regions, $i=1,...,n$ and $t=1,...,T$. We introduce several new methods of analysis to find a continuously varying offset between the multivariate time series $x_i(t)$ and $y_i(t)$. All four methods involve 7-day averaging; this is performed due to the consistent weekly patterns of COVID-19 reporting, with lower reporting on the weekends. Thus, let $\hat{x}_i(t)$ be the rolling 7-day case average, defined by
\begin{align}
    \hat{x}_i(t)=\frac17 \sum_{s=t-6}^t x_i(s), t=7,...,T,
\end{align}
and analogously let $\hat{y}_i(t)$ be the rolling 7-day death average. The following four methods, described in the proceeding subsections, are contributions to the literature.

\subsection{Probability vector method}
\label{sec:PDFmethod}

First, we estimate a continuously varying offset between multivariate time series $x_i(t)$ and $y_i(t)$ via a comparison of probability vectors of total counts. Let $p^X(t) \in \mathbb{R}^n$ be the probability vector of 7-day rolling averaged cases in each state, observed over an interval $[t-6,t]$. That is,
\begin{align}
\label{eq:PDF}
  p^X_i(t) = \frac{\hat{x}_i(t)}{ \sum_{j=1}^n \hat{x}_j(t)  } = \dfrac{\sum_{s=t-6}^t x_i(s)}{\sum_{s=t-6}^t \sum_{j=1}^{n} x_j(s)}, i={1},...,n, t=7,...,T.
\end{align}
Equivalently, eq. (\ref{eq:PDF}) shows that $p^X(t)$ is the probability vector of new cases in each state, observed across an interval $t-6\leq s \leq t$, divided by the total number of US cases across this period. Let $p^Y(t)$ be the analogous vector for deaths. As these probability vectors are suitably normalised, it is possible to compare them directly. Given two vectors $p,q \in \mathbb{R}^n$, let their $L^1$ distance be defined as $\|p-q\|_1= \sum_{i=1}^n |p_i - q_i|$.

Next, we define a \emph{search interval length} of $S=50$. With this, let the offset between the multivariate time series be defined by the following function:
\begin{align}
    f_1: [7, T-S] \cap \mathbb{Z} \to [0,S] \cap \mathbb{Z} ; \\
    t \mapsto \argmin_s \{ \| p^X(t) - p^Y(t+s) \|_1: s=0,1,...,S \}.
\end{align}
That is, for any probability vector of (averaged) cases at time $t$, $f_1(t)$ is defined as the time at most $S=50$ days in the future with the closest probability vector of (averaged) deaths. We remark that the domain of $f_1$ is restricted to $[7,T-S] \cap \mathbb{Z}$ to allow an entire search interval of $S=50$ days for each $t$. Were this not included, the function would be trivially bounded and decrease to zero as $t$ approached $T$.

\subsection{Affinity matrix method}
\label{sec:affinitymethod}

In this section, we estimate a continuously varying offset between multivariate time series $x_i(t)$ and $y_i(t)$ by comparing affinity matrices of counts between states. Let $D^X(t) \in \mathbb{R}^{n \times n}$ be the distance matrix between 7-day rolling averaged  cases in each state, observed over an interval $[t-6,t]$. That is,
\begin{align}
\label{eq:distancematrix}
  D^X_{ij}(t) = |\hat{x}_i(t) - \hat{x}_j(t)|, 
  i,j={1},...,n, t=7,...,T.
\end{align}
Let $D^Y(t)$ be the analogous matrix for deaths. Given such a distance matrix $D$, we associate a $n \times n$ affinity matrix $A$ by
\begin{align}
    A_{ij}= 1 - \frac{D_{ij}}{\max D}.
\end{align}
This is suitably normalised with all elements in $[0,1]$ to allow direct comparison between different affinity matrices. Let $A^X(t)$ and $A^Y(t)$ be the affinity matrices corresponding to the distance matrices $D^X(t)$ and $D^Y(t)$, respectively. Given two matrices $A,B \in \mathbb{R}^{n \times n}$, let their $L^1$ distance be defined as $\|A-B\|_1= \sum_{i,j=1}^n |A_{ij} - B_{ij}|$.

Again we use a \emph{search interval length} of $S=50$. With this, let the offset between the multivariate time series be defined by the following function:
\begin{align}
    f_2: [7, T-S] \cap \mathbb{Z} \to [0,S] \cap \mathbb{Z} ; \\
    t \mapsto \argmin_s \{ \| A^X(t) - A^Y(t+s) \|_1: s=0,1,...,S \}.
\end{align}
That is, for any affinity matrix between states' (averaged) cases at time $t$, $f_2(t)$ is defined as the time at most $S=50$ days in the future with the closest affinity matrix between (averaged) deaths. Again, the domain of $f_2$ is restricted to $[7,T-S] \cap \mathbb{Z}$ to allow a complete search interval of $S=50$ days. Were this not included, the function would be trivially bounded and decrease to zero as $t$ approached $T$.

\subsection{Inner product method}
\label{sec:innerproductmethod}

This section estimates a continuously varying offset between multivariate time series $x_i(t)$ and $y_i(t)$ via normalised inner products between individual states' time series. As before, we make use of the 7-day rolling averaged counts $\hat{x}_i(t)$ and $\hat{y}_i(t)$, but this time we restrict to one state at a time for our calculations. For the proceeding exposition, let $\hat{x}(t)$ and $\hat{y}(t)$ be the 7-day averaged counts of cases and deaths for a single candidate state.

Suppose $a\leq t \leq b$ and $c \leq t \leq d$ are two intervals within $[7,T]$ of equal length $L=b-a=d-c$. Let the normalised inner product between $\hat{x}(t)_{a \leq t \leq b}$ and $\hat{y}(t)_{c \leq t \leq d}$ be defined and notated as follows:
\begin{align}
\label{eq:normip}
    <\hat{x}(a:b),\hat{y}(c:d)>_n = \dfrac{ \sum_{t=0}^L \hat{x}(a+t) \hat{y}(c+t)}{ \left(\sum_{t=a}^b \hat{x}(t)^2 \right)^\frac12 \left(\sum_{t=c}^d \hat{y}(t)^2 \right)^\frac12 }.
\end{align}
This normalised inner product is derived from the standard Euclidean inner product on $\mathbb{R}^{L+1}$. Indeed, for $\mathbf{u},\mathbf{v} \in \mathbb{R}^{L+1}$, let $<\mathbf{u},\mathbf{v}>=\sum_{i=1}^{L+1} u_i v_i$. Then $<.,.>$ is symmetric, bilinear and positive-definite; $<\mathbf{u},\mathbf{u}>=\sum_{i=1}^{L+1} u_i^2$ recovers the Euclidean norm on $\mathbb{R}^{L+1}$. We can re-express eq. (\ref{eq:normip}) as follows:
\begin{align}
\label{eq:normip2}
    <\hat{x}(a:b),\hat{y}(c:d)>_n = \dfrac{ <\mathbf{u},\mathbf{v}> }{ \left(\ <\mathbf{u},\mathbf{u}> \right)^\frac12 \left( <\mathbf{v},\mathbf{v}> \right)^\frac12 },
\end{align}
where $\mathbf{u}=\hat{x}(t)_{a \leq t \leq b}$ and $\mathbf{v}=\hat{y}(t)_{c \leq t \leq d}$. That is, eq. (\ref{eq:normip}) presents a normalised analogue of the standard Euclidean inner product on $ \mathbb{R}^{L+1}$.

We have chosen these normalised inner products to have maximal value 1 if and only if there is a proportionality relation $\hat{y}(t)=k\hat{x}(t+\tau)$ for all $t=c,...,d$ for some constant $k>0$ and offset $\tau$. Indeed, we are seeking the offset in time where deaths are most closely proportional to cases. They are more suitable than other metrics, such as correlation or distance correlation  \cite{Szkely2007}. Correlation or distance correlation would each return maximal value 1 if $y=kx + b$ for an additional constant $b$, which is unsuitable.

Next, we use a rolling window of length $L=150$ days in which to compute a varying maximised offset. This longer window is chosen here in order to capture undulations in the time series, which are necessary for the inner product comparison to work well. Indeed, the inner product is maximised when local maxima and minima in cases are aligned with future local maxima and minima in deaths. Within each window, we again use a search interval length of $S=50$. Then, let the offset between the univariate time series $\hat{x}_i(t)$ and $\hat{y}_i(t)$ for each state $i$ be defined by the following function:
\begin{align}
    g_i: [7, T-L]  \cap \mathbb{Z}\to [0,S] \cap \mathbb{Z};\\
    t \mapsto \argmax_{\tau} \{ <\hat{x}_i(t:t+L-\tau),\hat{y}_i(t+\tau:t+L)>_n: \tau=0,1,...,S \}.
\end{align}
Effectively, this function considers the interval $[t,t+L]$ as fixed and selects an appropriate offset only by considering case and death counts within the interval. For that purpose, we must consistently truncate the case time series at the end, and the death time series at the beginning, hence the computation of the normalised inner product between $\hat{x}(s)_{t \leq s \leq t+L-\tau}$ and $\hat{y}(s)_{t+\tau \leq s \leq t+L}$.

Finally, the overall offset between the multivariate time series $\hat{x}_i(t)$ and $\hat{y}_i(t)$, $i=1,...,n$ is simply defined as
\begin{align}
    g: [7, T-L] \cap \mathbb{Z} \to \mathbb{R};\\
    g(t)=\frac{1}{n} \sum_{i=1}^n g_i(t).
    \label{eq:gaveraging}
\end{align}
Due to the averaging process, this is not necessarily integer-valued.

\subsection{Vector comparison method}
\label{sec:warping}

In this final methodological section, we estimate not only a continuously varying offset between multivariate time series of cases and deaths, but also a time-varying mortality rate. We proceed by directly comparing vectors of cases and deaths and attempting to minimise appropriate linear combinations thereof. We present multiple variations within this framework based on different ``loss''  functions - these record differences between vectors of cases and deaths, up to linear rescaling. As in Section \ref{sec:innerproductmethod}, we are seeking an offset in time where deaths are most closely proportional to cases. Unlike Section \ref{sec:innerproductmethod}, we use all states concurrently.

First, we define an $L^1$ 1-day loss function as follows:
\begin{align}
\label{eq:difference1}
  \mathcal{L}^1_{1}(t,\tau,\lambda)=  \sum_{i=1}^n |\hat{x}_i(t) - \lambda \hat{y}_i(t+\tau)|.
\end{align}
We remark that $\lambda$ plays the role of the inverse of the mortality rate between cases and deaths and is chosen for increased interpretability when plotting our results. Equivalently, we expect one death out of every $\lambda$ cases (for an optimal $\lambda$).

There are three parameters we can vary in our loss function. First, we can use sums of squares (an $L^2$ difference) rather than the above $L^1$ difference. Second, rather than fixing a single day $t$, we could compute a loss function over a longer period of length $P$. For example, we define an $L^1$ $P$-day loss function as follows:
\begin{align}
\label{eq:difference2}
   \mathcal{L}^1_{P}(t,\tau,\lambda)= \sum_{i=1}^n \sum_{j=0}^{P-1}|\hat{x}_i(t+j) - \lambda \hat{y}_i(t+j+\tau)|.
\end{align}
Third, we could modify the loss functions with a division term. For example, we define an $L^1$ 1-day divided loss function as follows:
\begin{align}
\label{eq:difference3}
    \mathcal{L}^1_{1,div}(t,\tau,\lambda)= \sum_{i=1}^n \frac{|\hat{x}_i(t) - \lambda \hat{y}_i(t+\tau)|}{|\hat{x}_i(t)|}.
\end{align}
We can also combine these modifications, for example using sums of squares in (\ref{eq:difference2}) and (\ref{eq:difference3}).

If we use $L^2$ differences, we have an analytically determined value of $\lambda$ that minimises the function for any candidate $\tau$. For example, consider the $L^2$ 1-day loss function,
\begin{align}
\label{eq:difference5}
     \mathcal{L}^2_{1}(t,\tau,\lambda) = \sum_{i=1}^n |\hat{x}_i(t) - \lambda \hat{y}_i(t+\tau)|^2.
\end{align}
The partial derivative with respect to $\lambda$ is 
\begin{align}
\label{eq:difference6}
    2\sum_{i=1}^n \hat{y}_i(t+\tau)(\lambda \hat{y}_i(t+\tau) - \hat{x}_i(t)).
\end{align}
By minimising a quadratic, there exists a distinguished value
\begin{align}
\label{eq:distinguishedlambda}
    \hat{\lambda} = \frac{\sum_{i=1}^n \hat{y}_i(t+\tau)\hat{x}_i(t) }{\sum_{i=1}^n  \hat{y}_i(t+\tau)^2 }
\end{align}
that minimises $\mathcal{L}^2_{1}(t,\tau,\lambda)$ for fixed $t$ and $\tau$, and similarly for other $L^2$ loss functions.

With the framework of loss functions as defined above, we can now define the continuous time-varying offset and associated inverse mortality. Again, we use a search interval length of $S=50$ days. For any candidate loss function $\mathcal{L}$, we define the following function:
\begin{align}
    \mathbf{h}_\mathcal{L}: [7,T-S] \cap \mathbb{Z} \to [0,S] \cap \mathbb{Z} \times \mathbb{R}^+ \subset \mathbb{R}^2 \\
    t \mapsto \argmin_{\tau, \lambda} \{\mathcal{L}(t,\tau,\lambda) : \tau=0,...,S, 1 \leq \lambda \leq 200 \}.
\end{align}
We remark that $\mathbf{h}_\mathcal{L}$ effectively has two outputs. We write $\mathbf{h}_\mathcal{L}(t)=(\tau_\mathcal{L}(t),\lambda_\mathcal{L}(t)) \in \mathbb{R}^2$. Then, $\tau_\mathcal{L}(t)$ gives the time-varying offset between the multivariate time series, while $\lambda_\mathcal{L}(t)$ gives the continuously varying inverse mortality rate. For the $L^2$ loss functions, $\lambda_\mathcal{L}(t)$ can be determined analytically through equation (\ref{eq:distinguishedlambda}). For the $L^1$ loss functions, the optimisation can be performed via a grid search. For the inverse mortality rate $\lambda$, we search over a closed bounded interval $[1,200]$, corresponding to a search of mortality rate between 0.5 and 100\%.

\section{Results}
\label{sec:results}

\begin{figure*}
    \centering
    \includegraphics[width=1.1\textwidth]{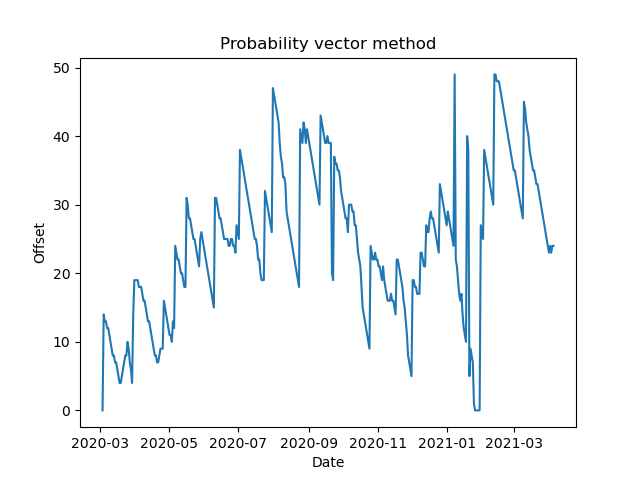}
    \caption{Continuous time-varying offset $f_1(t)$ determined by the probability vector method, detailed in Section \ref{sec:PDFmethod}. A pattern of increase, decrease and then increase is observed. In order to accommodate the $S=50$-day search window, the indexed dates end 50 days from the end of our analysis period.}
    \label{fig:PDF}
\end{figure*}

\begin{figure*}
    \centering
    \includegraphics[width=1.1\textwidth]{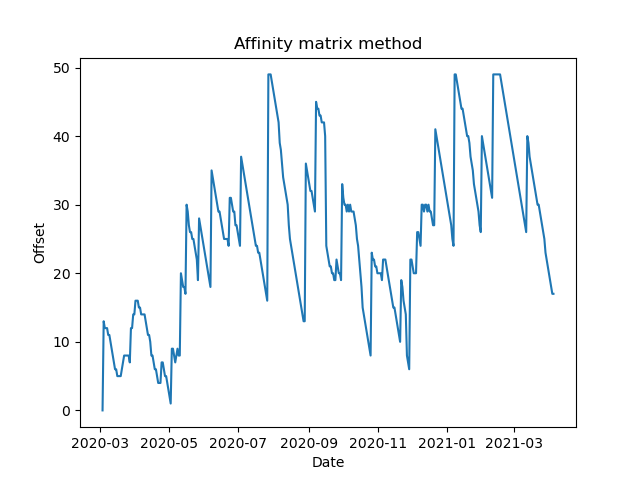}
    \caption{Continuous time-varying offset $f_2(t)$ determined by the affinity matrix method, detailed in Section \ref{sec:affinitymethod}. A pattern of increase, decrease and then increase is observed. In order to accommodate the $S=50$-day search window, the indexed dates end 50 days from the end of our analysis period.}
    \label{fig:affinity}
\end{figure*}

Figures \ref{fig:PDF} and \ref{fig:affinity} show the determined time-varying offset for the probability vector and affinity matrix method, respectively. In these plots, the value of the function at a date index of 2020-03, for example, records the optimal offset $\tau$ between cases at 1 March 2020 and deaths $\tau$ days later. Considerable similarity in these results is observed, which is to be expected, as both methods work similarly, by finding a future day in deaths with similar internal structure among states as a given day in cases. An ``up-down-up'' pattern is visible. Initially, the calculated offset during March 2020 is about 10 days. The offset rises to approximately 30 around September 2020 and then declines once more to 10-20 towards the end of 2020. Subsequently, an increase is observed to around 40, albeit with some irregularity during February 2021.

\begin{figure*}
    \centering
    \includegraphics[width=1.1\textwidth]{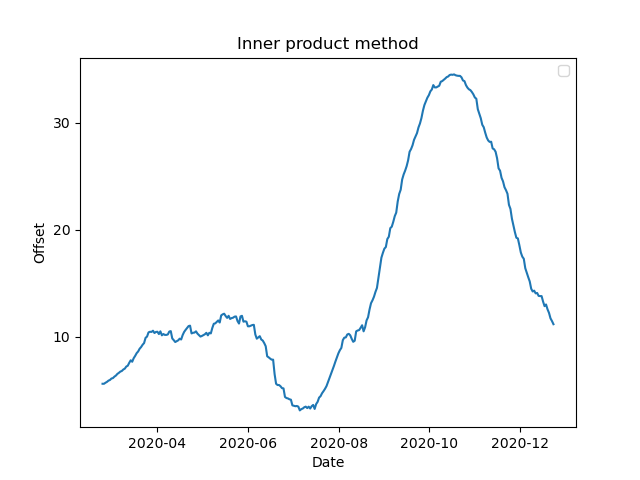}
    \caption{Continuous time-varying offset $g(t)$ determined by the normalised inner product method, detailed in Section \ref{sec:innerproductmethod}. A pattern of increase, decrease and then increase is observed. This function is substantially smoother than Figures \ref{fig:PDF} and \ref{fig:affinity} due to the averaging in its definition (\ref{eq:gaveraging}). In order to accommodate the $L=150$-day rolling computation window, the indexed dates end 150 days from the end of our analysis period.}
    \label{fig:IP}
\end{figure*}

Figure \ref{fig:IP} shows the offset for the normalised inner product method. This function is substantially smoother than the other offset functions in this paper due to the averaging in its definition (\ref{eq:gaveraging}). Again, an ``up-down-up'' pattern is observed but with consistently smaller values than the previous two methods. The determined offset rises from approximately 5 in March 2020 to 10, back down almost to zero, and up to a peak of over 30. We remark that the inner product method examines data $L=150$ days in advance, while the other methods search data only $S=50$ days in advance, so the determined offsets in Figure \ref{fig:IP} lead ahead of all the other figures. Aside from this, the inner product method is quite different to the other methods presented in this manuscript. Indeed, Section \ref{sec:innerproductmethod} shows how an offset is computed individually for each state, while every other method uses the entire multivariate data in conjunction.

\begin{figure*}
    \centering
    \begin{subfigure}[b]{0.65\textwidth}
        \includegraphics[width=\textwidth]{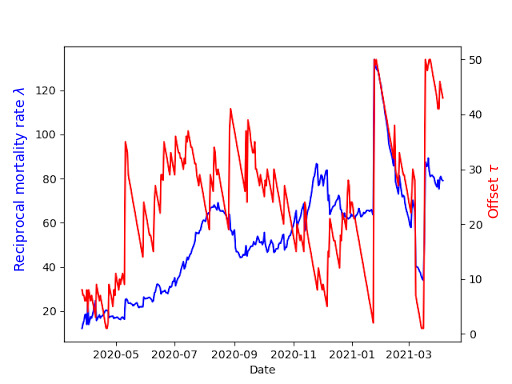}
        \caption{}
    \label{fig:L1_1day_divided}
    \end{subfigure}
    \begin{subfigure}[b]{0.65\textwidth}
        \includegraphics[width=\textwidth]{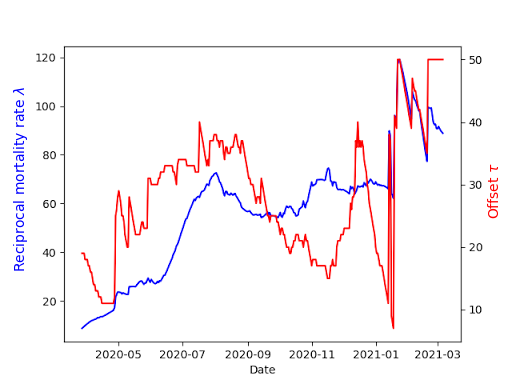}
        \caption{}
    \label{fig:L1_30day_notdivided}
    \end{subfigure}
    \begin{subfigure}[b]{0.65\textwidth}
        \includegraphics[width=\textwidth]{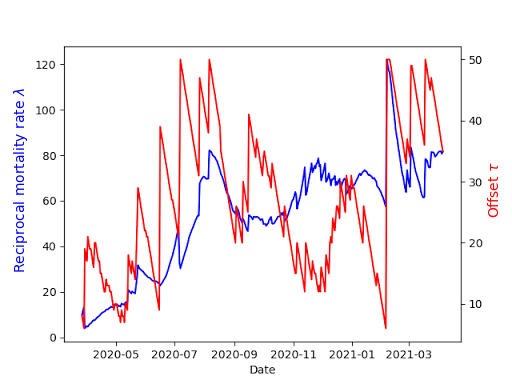}
        \caption{}
    \label{fig:L2_1day_notdivided}
    \end{subfigure}
    \caption{Several alternative time-varying offset functions computed within our framework of direct vector comparison, detailed in Section \ref{sec:warping}. To show a range of options, we present examples with every possible variation of parameters. In (a), we use an $L^1$ 1-day divided loss function. In (b), we use an $L^1$ 30-day loss function (without division). In (c), we use an $L^2$ 1-day loss function (without division), involving an analytically determined $\lambda$ (as in (\ref{eq:distinguishedlambda})). A pattern of increase, decrease and then increase is observed in the offset $\tau$, and a rather consistent increase in the reciprocal mortality rate $\lambda$.}
    \label{fig:warping}
\end{figure*}

In Figure \ref{fig:warping}, we present three plots for three different loss functions within our vector comparison framework of Section \ref{sec:warping}. We make sure to trial some variation of all three available parameters in our loss function. In Figures \ref{fig:L1_1day_divided}, \ref{fig:L1_30day_notdivided} and \ref{fig:L2_1day_notdivided}, we use an $L^1$ 1-day divided loss function, an $L^1$ 30-day undivided loss function, and an $L^2$ 1-day undivided loss function, respectively. In all three figures, we display both the time-varying offset $\tau$ and inverse mortality rate $\lambda$. These figures are quite consistent with Figures \ref{fig:PDF} and \ref{fig:affinity} in the offset. Initially, the offset is consistently about 10 days, rising to 30, declining to 10-20, and dramatically rising to 40-50. Like Figures \ref{fig:PDF} and \ref{fig:affinity}, some irregularity is observed during January-February 2021. All three figures show a general increase in $\lambda$, signifying a consistent decrease in the mortality of COVID-19, at least with respect to observed cases and deaths. However, a brief period in reduction in $\lambda$ is observed in the fall of 2020.

We remark that we could easily apply an averaging or smoothing procedure to the probability vector, affinity matrix or vector comparison methods (Figures \ref{fig:PDF}, \ref{fig:affinity} and \ref{fig:warping}) to generate smoother curves like Figure \ref{fig:IP}, but have chosen to display the initial raw result. In addition, smoothing could be applied for use in a predictive setting.

\clearpage

\section{Discussion}
\label{sec:discussion}

All four methods and six plots displayed thus far exhibit an ``up-down-up'' pattern in the estimated time-varying offset between case and death time series. Early on, a small offset is observed - this has several explanations. First, US states were slow to implement effective and wide-scale testing regimes \cite{nytslowtesting}, so cases were likely substantially underreported. Secondly, treatments were limited, thus infected patients may have passed away from infection within a quicker time frame. Third, due to the novelty of the virus, many vulnerable individuals such as the elderly may have contracted the disease early on and passed away relatively quickly. Later on, it is likely that vulnerable individuals took greater precautions than the rest of the population.

Subsequently, the offset increases until July-September, depending on the precise method. (The offset estimated by the inner product method (Figure \ref{fig:IP}) peaks $\sim3$ months earlier, likely due to examining data $L=150$ days in advance rather than $S=50$ days in advance.) This could be attributed to improved treatment \cite{Remdesivir,Bloch2020,toczilizumab,Cao2020}, non-pharmaceutical interventions, including social distancing, business closures, and better management of nursing homes, and more widespread testing.

\begin{figure*}
    \centering
    \includegraphics[width=\textwidth]{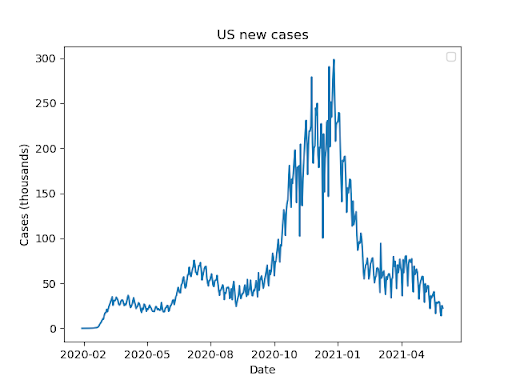}
    \caption{New daily cases for the entire United States.}
    \label{fig:totalUS}
\end{figure*}

Curiously, all methods observe a subsequent decrease in the estimated progression between cases and deaths. This decrease begins around August-September 2020 for the probability vector, affinity matrix and vector comparison methods (Figures \ref{fig:PDF}, \ref{fig:affinity} and \ref{fig:warping} respectively), and proportionately earlier for the inner product method. This period heralds a consistent worsening in the status of the pandemic throughout the US. As seen in Figure \ref{fig:totalUS}, cases consistently rise from early September to the end of 2020. In addition, many states relax and do not reimpose lockdown measures during this time \cite{wapo_allreopen}, and the colder climate yields worse outcomes both in terms of spread and illness \cite{Mallapaty2020}. The change in offset is not necessarily only due to individual progressions from infection to death, but involves mediating factors like stresses on hospital capacity. For example, perhaps initial waves of patients can be treated with ventilators, but these may quickly run out, causing more deaths from later cases.

The status of the pandemic changes drastically following the beginning of 2021. First, cases precipitously fall (Figure \ref{fig:totalUS}), perhaps following the increased gathering of people over Thanksgiving and Christmas. Second, the rollout of vaccines produced at the end of 2020 \cite{Polack2020,Walsh2020} targeted vulnerable populations first and had a beneficial effect on the mortality of COVID-19 among the elderly. The drastic change in the status of the pandemic during this time could be the cause of the irregularity observed in several figures. The dramatically higher determined offset at the end of the time window, at least for Figures \ref{fig:PDF}, \ref{fig:affinity} and \ref{fig:warping}, is a welcome testament to the effectiveness of vaccines and the still improving treatment for unvaccinated individuals.

One strength in our paper is the fact that four different methods, including different loss functions within the vector comparison framework, yield relatively similar results. The loss functions in Section \ref{sec:warping} allow variation of three parameters, which are all trialled at least once in the subfigures of Figure \ref{fig:warping}. In particular, the choice of whether to implement divided loss functions, such as in (\ref{eq:difference3}), notably changes the properties of the loss function. In an undivided loss function such as (\ref{eq:difference1}), larger states with larger absolute values of $\hat{x}_i(t)$ and $\hat{y}_i(t)$ are likely to disproportionately influence the selection of $\tau$ and $\lambda$. In a divided loss function, this is no longer the case. It is a strength of the framework of Section \ref{sec:warping} that this normalisation produces little difference in results.

Several limitations exist in this paper and even in future work. First, the framework of Section \ref{sec:warping} implicitly assumes (and aims to find) a constant (inverse) mortality rate $\lambda$ among all the states. While the US mostly has a similar standard of living and healthcare system quality from state to state, this is not uniformly the case, and different states differ substantially in population density and socioeconomic demographics. However, we believe that the remarks above, wherein two drastically different methods that prioritise larger states and all states, respectively, give similar results, show that perhaps this limitation is not too grave.

Second, it is notable that Figure \ref{fig:IP}, defined by the inner product method of Section \ref{sec:innerproductmethod}, is the one figure most different to the others. That is, it appears to be the odd one out relative to Figures \ref{fig:PDF}, \ref{fig:affinity}, \ref{fig:L1_1day_divided}, \ref{fig:L1_30day_notdivided} and \ref{fig:L2_1day_notdivided}. This difference is to be partially expected, as the inner product method works quite differently to the other methods. Namely, it examines $L=150$ days in advance rather than $S=50$ days, and computes an offset for each individual state rather than the multivariate time series as a whole. One may consider the outlier of Figure \ref{fig:IP} to be both a strength and limitation of the manuscript. It is potentially a limitation as it does not match the other figures exactly, but it may be a strength as it suggests that computing a separate offset for each state and simply averaging them is too naive a procedure. We remark that the inner product method is the simplest and the most closely related to (quite naive) existing methods of computing offsets between time series, such as cross-correlation \cite{Crosscorrelation}. In Section \ref{sec:innerproductmethod}, we explained why our chosen normalised inner product is more suitable in this context than an offset correlation (essentially equivalent to the method of cross-correlations). That is, the outlier status of Figure \ref{fig:IP} may have a notable upshot: that a full consideration of the multivariate structure of the time series is necessary, and not simply an individual consideration of each state at a time.

Third, extending our analysis into the future may be difficult due to complexities in the epidemiology of COVID-19 and the availability of data. Specifically, the rollout of vaccines has created very different progressions from cases to deaths in the vaccinated vs unvaccinated populations. In addition, the Centers for Disease Control and Prevention (CDC) has changed its reporting of cases among the vaccinated population, only tracking ``breakthrough cases'' that result in hospitalisation or death \cite{atlanticbreakthrough}. Future work could use the analysis presented in this paper, but more precise data needs to be collected and made available on an ongoing basis. More broadly, we encourage future work to carefully separate out the mathematical epidemiology of COVID-19 between vaccinated and unvaccinated populations, studying phenomena not limited to the offset between cases and deaths, and further exploring the positive impact of COVID-19 vaccines on the community. For example, future work could separate out the progression from COVID-19 infection to either death or recovery among the vaccinated and unvaccinated populations, including a consideration of ``long Covid'' \cite{Mahase2020}. There may be numerous non-trivial benefits to be discovered with careful analysis. At the same time, as near-entire vaccination of the US population seems unlikely (that is, the US is unlikely to reach herd immunity), measures to contain and reduce the impact of the virus on the healthcare system remain highly relevant for the reduction of casualties and economic and other social consequences \cite{Priesemann2021,Momtazmanesh2020}.

\section{Conclusion}
Overall, we have proposed four methods to determine a continuously varying offset between two multivariate time series and applied this to the state-by-state counts of COVID-19 cases and deaths in the United States. Our final method is a framework of loss functions in which we have trialled the variation of several parameters. Our methods exhibit considerable robustness with broadly similar results obtained, including under relatively substantial changes such as normalising by case counts in Section \ref{sec:warping} to de-prioritise larger states. Our findings reveal new insights into the time-varying progression from cases to deaths in the US and discuss how this reflects the changing status of the pandemic. We show that the estimated offset between cases and deaths rises between the first and second waves of COVID-19 in the US, falls towards the end of 2020, and dramatically rises in 2021. Minor modifications such as smoothing, combined with updated and reliable data on cases among the vaccinated and unvaccinated populations could provide a valuable predictive tool regarding future periods of high load on the healthcare system. Our analysis could also be applied to other multivariate time series outside epidemiology.

\section*{Data availability}

The dataset analysed during the current study is available at \cite{datasourcenyt}.

\section*{Acknowledgements}

The authors would like to thank Howard Bondell for helpful discussions.

\bibliographystyle{epj}
\bibliography{__references}
\end{document}